\documentclass[onecolumn,11pt]{IEEEtran}

\usepackage{color}
\usepackage{amsmath}
\usepackage{amssymb}
\usepackage{setspace}
\usepackage{bm}
\usepackage[final]{graphicx}

\newcommand{\dtilde}[1]{\tilde{\tilde{#1}}}

\newcommand{\Nr}[1]{{\mathcal N} \left(#1 \right)}
\newcommand{\E}[1]{E\left\{ #1 \right\}}

\newcommand{\N}[1]{{\mathcal N}\left(#1 \right)}

\renewcommand{\Pr}{{\rm Pr}}
\newcommand{\zev}{{\bf 0}}

\newcommand{\sigman}{\sigma_n^2}

\newcommand{\bartilde}[1]{\bar{\tilde{#1}}}

\newcommand{\Lcal}{{\mathcal L}}

\newcommand{\bv}{{\bf b}}

\newcommand{\Gv}{{\bf G}}
\newcommand{\Hv}{{\bf H}}
\newcommand{\hv}{{\bf h}}
\newcommand{\Iv}{{\bf I}}

\newcommand{\nv}{{\bf n}}

\newcommand{\rv}{{\bf r}}

\newcommand{\sv}{{\bf s}}
\newcommand{\Sv}{{\bf S}}

\newcommand{\defi}{\stackrel{\triangle}{=}}

\renewcommand{\(}{\left(}
\renewcommand{\)}{\right)}

\newcommand{\mtA}{{\mathcal{A}}}
\newcommand{\mtK}{{\mathcal{K}}}
\newcommand{\mtF}{{\mathcal{F}}}

\renewcommand{\baselinestretch}{1.72}

\title{Iterative (``Turbo'') Multiuser Detectors for Impulse Radio Systems}

\author{Eran Fishler$^\dag$, Sinan Gezici$^{*}$, and H. Vincent Poor$^\ddag$
\thanks{This research was supported in part by the National Science
Foundation under Grants ANI-03-38807 and CNS-06-25637.}
\thanks{$^\dag$ Department of Computer Science, New York University, NY 10012, USA, {\tt e-mail:
fishler@cs.nyu.edu}}
\thanks{$^{*}$ Department of Electrical and Electronics
Engineering, Bilkent University, Bilkent, Ankara 06800, Turkey, Tel:
+90 (312) 290-3139, Fax: +90 (312) 266-4192, {\tt e-mail:
gezici@ee.bilkent.edu.tr}}
\thanks{$^\ddag$ Department of Electrical Engineering, Princeton University, Princeton 08544, USA, Tel: (609) 258-2260, Fax: (609) 258-7305, {\tt e-mail: poor@princeton.edu}} }

\begin{document}

\maketitle

\begin{abstract}

In recent years, there has been a growing interest in multiple
access communication systems that spread their transmitted energy
over very large bandwidths. These systems, which are referred to as
ultra wide-band (UWB) systems, have various advantages over
narrow-band and conventional wide-band systems. The importance of
multiuser detection for achieving high data or low bit error rates
in these systems has already been established in several studies.
This paper presents iterative (``turbo'') multiuser detection for
impulse radio (IR) UWB systems over multipath channels. While this
approach is demonstrated for UWB signals, it can also be used in
other systems that use similar types of signaling. When applied to
the type of signals used by UWB systems, the complexity of the
proposed detector can be quite low. Also, two very low complexity
implementations of the iterative multiuser detection scheme are
proposed based on Gaussian approximation and soft interference
cancellation. The performance of these detectors is assessed using
simulations that demonstrate their favorable properties.

\textit{Index Terms---}$\,$Ultra wide-band (UWB), impulse radio
(IR), iterative multiuser detection, soft interference cancellation.

\end{abstract}

\newpage

\section{Introduction}

In recent years, there has been a growing interest in ultra
wide-band (UWB) systems, which resulted in the U.S.
Federal Communications Commission (FCC) regulations that allow,
under several restrictions, the widespread use of such systems. The
common definition of UWB systems, which was adopted by the FCC as
well, states that a system is a UWB system if both the absolute and
the fractional bandwidths are large. The absolute bandwidth should
be at least $0.5$ GHz, while the fractional bandwidth, which is the
signal bandwidth divided by the carrier frequency, is at least
$20\%$ \cite{FCC:02}. UWB systems offer many advantages over
narrow-band or conventional wide-band systems. Among these
advantages are reduced fading margins, simple transceiver designs,
low probability of detection, good anti-jam capabilities, and
accurate positioning (see, \cite{Choi:02,Win:98,Gezici05mag}, and
references therein). The advantages of UWB technology have caused
this technology to be considered for use as the physical layer of
several applications; for example, the IEEE 802.15.4a wireless
personal area network (WPAN) standard employs this technology as one
of the signaling options \cite{SG4}.

There are many signaling methods for transmitting over UWB channels,
and it is obvious that, apart from engineering difficulties, one can
use any existing spread spectrum technique for transmitting over UWB
channels \cite{Foerster:02,Welborn:02}. However, these difficulties
might be quite significant, preventing the actual use of
conventional spread-spectrum methods for transmitting over UWB
channels. Consider, as an example, long-code direct-sequence
code-division-multiple-access (DS-CDMA) systems. In these systems,
implementing even the simplest detector, namely the matched filter
detector, requires sampling of the received signal at least at the
chip rate, which under the current regulations might be as large as
$7.5$ GHz. Such sampling rates are difficult to achieve, and result
in high power consumption.

In order to overcome some of the difficulties associated with UWB
signaling, impulse radio (IR) systems, and especially time-hopping
impulse radio (TH-IR) systems have been proposed as the preferred
modulation scheme for UWB systems \cite{Scholtz:93}. In TH-IR
systems, a train of short pulses is transmitted, and the information
is usually conveyed by either the polarity or location of the
transmitted pulses. In addition, in order to allow many users to
share the same channel, an additional random (or pseudo-random) time
shift, known to the receiver, is added to the starting point of each
pulse. This way, probability of catastrophic collisions between two
users transmitting over the same channel at the same time is
significantly reduced \cite{Scholtz:93}.

TH-IR modulation, e.g., binary phase shift keyed (BPSK) TH-IR, to be discussed in the following sections, has many advantages over conventional modulation techniques. By using very short pulses, the transmitted energy is spread over a very large bandwidth. In addition, by using pseudo-random time intervals between the transmitted pulses and random pulse polarities, spectral lines and other spectral
impairments are avoided \cite{sinanMP}. The implementation of the
receiver is usually easier for this technique because the channel is
excited for only a fraction of the total transmission time. For
example, the matched filter detector needs to sample the filter
matched to the received pulse only at time instants when pulses
corresponding to the user of interest arrive at the receiver.
Moreover, base-band pulses are typically used in UWB systems, saving
the need for complex frequency synchronization and
tracking\footnote{It should be noted, however, that if the channel
is composed of a very large number of equipower paths, then the
receiver complexity becomes very large due to the need to sample all
of them in order to achieve diversity combining.}. These advantages
make TH-IR the preferred modulation scheme for transmitting over UWB
channels in various applications. It should be noted that 
IR-UWB has been chosen as one of the modulation formats for the IEEE
802.15.4a WPAN standard.

It has been observed \cite{FishlerMUD:04,Li:02a,Scholtz:01,Yang:02b}
that the transmitted and received signals of TH-IR systems can be
described by the same models used for describing the transmitted and
received signals of DS-CDMA systems. The main difference between
classical DS-CDMA signals and TH-IR signals is that TH-IR signals
use spreading sequences whose elements belong to the ternary
alphabet, i.e., $\{-1,0,+1\}$, instead of the binary alphabet, i.e.,
$\{-1,+1\}$. This observation leads to the immediate conclusion that
every multiuser detector designed for CDMA systems can be used in
TH-IR systems as well. In particular, the optimal multiuser detector
can be easily deduced from \cite{Verdu:98}, and the complexity of
this detector for systems transmitting over multipath channels is
known to be exponential in the number of active users and the number
of transmitted symbols falling within the delay spread of the channel. Linear receivers can be designed as well,
resulting in multiuser detectors having complexity that is
polynomial in the number of active users and the size of the
observation windows used by the detector \cite{Buzzi:01,Li:02}.

Although the classical algorithms for multiuser detection can be
used in TH-IR systems, it is evident that low complexity multiuser
detection algorithms for systems that use generalized spreading
sequences in general and IR systems in particular are required.
These detectors should exploit the special type of signals TH-IR
systems transmit in order to reduce the complexity of multiuser
detectors. In \cite{FishlerMUD:04}, an iterative multiuser detector
exploiting the special structure of TH-IR signals is proposed for
additive white Gaussian noise (AWGN) channels. Iterative multiuser detectors can be designed for TH-IR systems by considering the TH-IR signaling structure as a concatenated coding system, where the inner code is the modulation and the outer code is the repetition code. Such a technique makes use of the similarity between TH-IR signaling and bit interleaved coded modulation (BICM), where the inner code is modulation and the outer code is channel coding \cite{BCM}, \cite{Vandendorpe:04}, \cite{Vesel:06}, \cite{Vesel:05}.

In this paper, we first present an extension of the iterative
multiuser detector in \cite{FishlerMUD:04} to more realistic
multipath channels. Namely, we propose an iterative detector
structure that combines energy from a number of multipath
components. Although only random TH-IR systems are described in the
sequel, the multiuser detectors presented in this paper can be
applied to any other type of DS-CDMA system whose spreading
sequences contain large fraction of zeros. As such the contribution of this paper goes beyond the theory of UWB systems into the theory of general DS-CDMA systems. In addition, we propose two very low-complexity implementations of the iterative algorithm, which are based on Gaussian approximation for weak interferers, and on soft
interference cancellation.

The rest of the paper is organized as follows: In Section II, the
signal model that is used throughout the paper is described. In
Section III, an iterative multiuser detector, called the
pulse-symbol iterative detector, is presented for
frequency-selective environments. Then, two novel and low-complexity
implementations of the proposed receiver are described in Section
IV. In Section V, simulations demonstrating the performance of the
proposed detector when transmitting over indoor UWB channels are
presented. Finally, a summary and some concluding remarks are provided in Section VI.

\section{Discrete-Time Signal Model}

TH-IR systems can be modeled as DS-CDMA systems with generalized
spreading sequences that take values from the set $\{-1,0,+1\}$
\cite{Martret:00a,GeziciTCOM}. Therefore, a $K$-user DS-CDMA
synchronous system transmitting over a frequency-selective channel
is considered in order to obtain the discrete-time signal model for
a TH-IR system\footnote{The synchronous assumption is made for
notational convenience, but as we discuss in the sequel, the
proposed algorithm works equally well in asynchronous systems.}. It is assumed that each user transmits a packet of $P$
information symbols, and $N$ denotes the processing gain of the
system. In addition, the channel between each user and the receiver
is modeled to have $L$ taps, and $\hv_k = [h_1^k\cdots h_L^k]$
denotes the discrete time channel impulse response between the $k$th
transmitter and the receiver. Finally, $\sv_{k,i}=[s^k_{i,0}\cdots
s^k_{i,N-1}]$ represents the spreading sequence that the $k$th user
uses for spreading its $i$th information symbol. Note that if
$\sv_{k,i} = \sv_{k,j}$ for every $i$ and $j$, then the systems is
a short-code system; otherwise it is a long-code system.

A chip-sampled discrete-time model for the received signal can be
described by the following model:
   \begin{gather}
      \rv = \sum_{k=1}^K \sqrt{E_k}\,\Hv_k\Sv_k\bv_k +\nv,
   \end{gather}
where, for the $k$th user ($k=1,\ldots,K$): $E_k$ is the transmitted
energy per symbol; $\Hv_k$ is an $(NP+L-1)\times NP$ matrix, whose
$i$th column is equal to $[{\bf 0}_{i-1}, \hv_k, {\bf 0}_{NP-i}]^T$
and ${\bf{0}}_l$ is the all zero row vector of length $l$; $\Sv_k$
is an $NP\times P$ spreading matrix containing the $P$ spreading
sequences that the $k$th user uses for spreading the transmitted
symbols,
$\Sv_k=\left[[\sv_{k,1}\,\zev_{N(P-1)}]^T,[\zev_N\,\sv_{k,2}\,\zev_{N(P-2)}]^T,
\ldots, [\zev_{N(P-1)}\,\sv_{k,P}]^T\right]$; and $\bv_k=[b_1,
\ldots, b_P]^T$ is the vector containing the transmitted information
symbols of the $k$th user. Throughout this paper, it is assumed that
the transmitted information symbols are binary (i.e., elements of
$\{-1,+1\}$) although the extension to more general cases is
straightforward. Here, $\nv=[n_1, \ldots, n_{NP+L-1}]^T$ is the
sampled additive noise vector, assumed to be normally distributed
with zero mean and correlation matrix $\sigman\Iv$, i.e., $\nv \sim
\Nr{0,\sigman\Iv}$. In the sequel, this system is referred to as a
BPSK TH-IR system.

Denote by $\bv \defi [\bv_1^T,\bv_2^T,\ldots, \bv_K^T]^T$ the vector
containing the transmitted symbols of the various users, by $\Sv$
the block diagonal matrix with the users' spreading matrices on its
diagonal, and by $\Hv \defi [\Hv_1, \Hv_2, \ldots, \Hv_K]$ the
concatenation of the users' channel matrices. With the aid of
$\Hv,\Sv,$ and $\bv$, the following model for the received signal
can be deduced:
   \begin{gather}
      \rv = \Hv\Sv\bv + \nv.
      \label{e. final model}
   \end{gather}
In deriving (\ref{e. final model}), it is assumed without loss of
generality that the users' channel impulse responses are scaled to
absorb the transmitted energy per bit.

Equation (\ref{e. final model}) can also be used to describe DS-CDMA
systems, in which case it is usually assumed that all the elements
of $\Sv$ belong to $\left\{ \pm \frac{1}{\sqrt{N}} \right\}$, where
$N$ is the spreading gain. IR systems are, in a sense,
generalizations of DS-CDMA systems, where in IR systems all the
elements of $\Sv$ belong to $\left\{\pm \frac{1}{\sqrt{N_f}}, 0
\right\}$, where $N_f$ is the number of pulses (or ``chips" in the
CDMA terminology) each user transmits per information symbol. Since
each symbol interval in an IR system is divided into $N_f$ equal
intervals, called \textit{frame}s, and a single pulse is transmitted
in each frame, $N_f$ is also called the number of frames per symbol.

In practice each user, say the $k$th user, is assigned a random, or
a long pseudo-random, TH sequence, denoted by $\{c_j^k\}$. This
sequence is known to the receiver, but the elements of this sequence
can be modeled for analytical purposes as independent and
identically distributed (i.i.d.) random variables, uniformly
distributed in $\{0,1, \ldots, N_c-1\}$. Denote by
$\sv_k=[\sv_{k,1}^T,\sv_{k,2}^T, \ldots, \sv_{k,P}^T]$ the
concatenation of the spreading sequences of the $k$th user. The
elements of $\sv_k$ are related to the $k$th user's TH sequence as
follows: the elements of $\sv_k$ corresponding to indices
$\{(j-1)N_c+c_j^k+1\}_{j=1}^{N_f P}$ are binary random variables,
while the remaining elements are zero. Note that random CDMA systems
can be described by this model by taking $N_f=N$.

\section{The Pulse-Symbol Iterative Detector}

In this section, a low-complexity receiver structure, called the
``pulse-symbol (iterative) detector'' is proposed for TH-IR systems
in frequency selective environments. Since the receiver does not
require chip-rate or Nyquist rate sampling, it facilitates simple
implementations in the context of UWB systems.

Denote by $\Lcal^k=\{l_1^k,\ldots,l_M^k\}$, with
$l_m^k\in\{1,2,\ldots,L\}$ and $M\leq L$, the indices of the signal
paths the receiver combines for user $k$. In other words, the
proposed receiver samples the received signal at the time instances
when pulses arrive through the paths indexed by $\Lcal^k$ for
$k=1,\ldots,K$. It can be easily seen that these sampling times are
$\{((j-1)N_c+c_j^k +l_m^k )T_c \}_{j=1,k=1,m=1}^{N_fP,\,K,\,M}$,
where $T_c$ is the pulse width. Denote by $r_{j,m}^k$ the received
sample corresponding to the $j$th pulse of the $k$th user via the
$m$th signal path. Note that the total number of samples per symbol
from all frames and signal paths of all users can be as high as
$N_fMK$, which can result in a very high-complexity receiver
structure. Therefore, we consider a receiver that combines the
samples from different multipath components in each frame by maximal
ratio combining (MRC) for each user. Let $\tilde{r}_j^k$ denote this
combined sample in the $j$th frame of user $k$. Then,
\begin{gather}\label{e. combSample}
\tilde{r}_j^k=\sum_{m=1}^{M}h_{l_m^k}^kr_{j,m}^k,
\end{gather}
and the samples from user $k$ can be expressed as
$\tilde{\rv}_k=[\tilde{r}_1^k\cdots \tilde{r}_{N_fP}^k]$. The
proposed receiver is depicted in Figure \ref{f: basic receiver}. It
is easy to verify that $r_{j,m}^k$ is the $((j-1)N_c+c_j^k+l_m^k)$th
element of $\rv$ defined in (\ref{e. final model}), and therefore a
matrix, $\Gv_k$, which performs selection and MRC of selected
samples, can be designed such that $\tilde{\rv}_k = \Gv_k\rv$.

Based on the samples obtained as in (\ref{e. combSample}), the
pulse-symbol detector performs an iterative estimation of users'
symbols. In general, iterative algorithms provide low complexity and
close-to-optimal solutions for many problems (see,
\cite{Hagenauer:96,Poor:01,Wang:99}, \cite{Vandendorpe:04},
\cite{Vesel:06}, among many others; a review is found in
\cite{Poor:02}). The main property of the problems that can be
solved efficiently by iterative techniques is that these problems
have a very special structure, which allows productive use of
iterative procedures. Consider as an example the problem of joint
multiuser detection and decoding of error correcting codes in CDMA
systems \cite{Poor:01}. In this problem, one can employ any
multiuser detection algorithm (or more precisely a multiuser
receiver \cite{Tse:99}) that results in soft decision statistics
about every channel symbol. These soft decisions can be fed into any
soft decoding algorithm, and the result will be the estimated
information symbol. Turbo based algorithms provide an efficient way
of iterating between the results obtained by the two constituent
algorithms, where each one of these algorithms is designed to solve
one part of the problem. Although no such structure exists in the
problem of multiuser detection of TH-IR signals, some of the {\em{a
priori}} information can be neglected in order to impose a structure
suitable for an iterative decoding algorithm. In other words, the
spreading operation is regarded as a simple error correcting
encoding to facilitate iterative solutions. In this light, TH-IR
signaling can be considered as a concatenated coding system, where
the inner code involves the modulation of a UWB pulse, and the outer
code is a repetition code\footnote{Unlike conventional turbo receivers, there is not a separate interleaver unit between the coding units in the proposed structure. However, the function of an interleaver in reducing the correlation between the soft output of each decoder unit and the input data sequence (called the iterative decoding suitability  criterion \cite{IDS1}, \cite{IDS2}) is performed by the TH and polarity randomization codes in the proposed system. By means of TH and polarity codes \cite{sinanTSP}, inputs to the demodulator and the decoder blocks become essentially independent.}. This structure is similar to BICM, for which modulation and channel coding comprise the inner and outer codes, respectively \cite{BCM}, \cite{Vandendorpe:04}.

Consideration of TH-IR systems as BICM systems facilitates the
design of the pulse-symbol iterative detector, which is composed of
two stages \cite{FishlerMUD:04}. The first stage is denoted as the
``pulse detector'', while the second stage is denoted as the
``symbol detector'', and the detector iterates between these stages.
In the first stage, it is assumed that different pulses from the
same user correspond to independent information symbols, while in
the second stage the information that several pulses from the same
user correspond to the same information symbols is exploited. The
second stage acts effectively as a decoder.

\subsection{The Pulse Detector}

Denote by $b_j^k$ the information symbol carried by the $j$th pulse
of the $k$th user. Note that although we know {\em a priori} that
$b_{(i-1)N_f+1}^k = \cdots = b_{iN_f}^k$ for every $k=1,\ldots,K$
and $i=1,\ldots, P$, this information will be ignored by the pulse
detector. As such, at the $n$th iteration the pulse detector
computes the {\em a posteriori} log-likelihood ratio (LLR) of
$b_j^k$, given $\tilde{r}_j^k$ in (\ref{e. combSample}), the
information about the transmitted pulses from other users and the
{\em a priori} information about $b_j^k$ provided by the symbol
detector, as
   \begin{gather}
      L_1^n(b_j^k ) \defi \log \frac{ \Pr( b_j^k = 1 | \tilde{r}_j^k ) } {
      \Pr\( b_j^k = -1 | \tilde{r}_j^k \) } = \log \frac{ f\(
      \tilde{r}_j^k | b_j^k = 1 \) }{f\( \tilde{r}_j^k| b_j^k=-1\)} +
      \log \frac{\Pr\( b_j^k = 1 \) }{\Pr\( b_j^k=-1 \) },
      \label{e. P-D L_1}
   \end{gather}
for $j=1,\ldots,PN_f$ and $k=1,\ldots,K$, where $f\( \tilde{r}_j^k |
b_j^k = i \) $ is the likelihood of the $j$th combined sample
corresponding to the $k$th user given that the transmitted symbol
was $i \in \pm 1$. It is seen that the {\em a posteriori} LLR is the
sum of the {\em a priori} LLR of the transmitted symbol, $\log
\frac{\Pr\( b_j^k = 1 \) }{\Pr\( b_j^k=-1 \)} \defi
\lambda_2^{n-1}(b_j^k)$, and the {\em extrinsic} information
provided by the pulse detector about the transmitted symbol, $\log
\frac{ f\( \tilde{r}_j^k | b_j^k = 1 \) }{f\( \tilde{r}_j^k |
b_j^k=-1\)}\defi \lambda_1^n(b_j^k)$ \cite{FishlerMUD:04}.

We first consider the computation of $\log f\( \tilde{r}_j^k | b_j^k
\)$ in (\ref{e. P-D L_1}). From (\ref{e. final model}), it is easy
to deduce the following model for $r_{j,m}^k$, which is the received
sample from the $m$th path of the $k$th user's signal in the $j$th
frame:
   \begin{gather}
      r_{j,m}^k = [\Hv]_{l(j,k,m):}\Sv\bv + n_{l(j,k,m)}
      = \sum_{\tilde{q}=1}^K\sum_{\tilde{a}=0}^{N_fP-1}b_{\lfloor
      {\tilde{a}}/N_f \rfloor}^{\tilde{q}}[\Sv_{\tilde{q}}]_{{\tilde{a}}N_c+c_{\tilde{a}}^{\tilde{q}},
      \lfloor {\tilde{a}}/N_f \rfloor}h_{l(j,k,m)-{\tilde{a}}N_c-c_{\tilde{a}}^{\tilde{q}}}^{\tilde{q}}
      + n_{l(j,k,m)},
      \label{e. pulse model - complicated}
   \end{gather}
where $l(j,k,m)$ is the arrival time of the $j$th pulse of the $k$th
user via the $m$th path, that is $l(j,k,m) = (j-1)N_c+c_j^k+l_m^k$;
$[\Hv]_{l(j,k,m):}$ is the $l(j,k,m)$th row of $\Hv$;
$[\Sv_m]_{k,l}$ is the $(k,l)$th element of the matrix $\Sv_m$; and
$n_{l(j,k,m)}$ is the $l(j,k,m)$th element of the noise vector,
$\nv$. This model can be simplified further by noting that the vast
majority of the summands in (\ref{e. pulse model - complicated}) are
zero. Let $\mtA$ denote the set of distinctive
$({\tilde{q}},{\tilde{a}})$ pairs in the right-hand-side (RHS) of
(\ref{e. pulse  model - complicated}) such that the corresponding
element in the double sum is not zero; i.e.\footnote{Note that the
dependence of $\mtA$ on $j$, $k$ and $m$ is not shown explicitly for
notational simplicity.},
\begin{gather}\label{eq:setA}
\mtA=\{(\tilde{q},\tilde{a})\in\mtK\times\mtF\,\,|\,\,\,[\Sv_{\tilde{q}}]_{{\tilde{a}}N_c+c_{\tilde{a}}^{\tilde{q}},
      \lfloor {\tilde{a}}/N_f \rfloor}h_{l(j,k,m)-{\tilde{a}}N_c-c_{\tilde{a}}^{\tilde{q}}}^{\tilde{q}}\ne0\},
\end{gather}
where $\mtK=\{1,\ldots,K\}$ and $\mtF=\{0,\ldots,PN_f-1\}$. If
$K_{j,m}^k$ represents the number of summands in (\ref{e. pulse
model - complicated}) that are different from zero, $\mtA$ consists
of $K_{j,m}^k$ pairs. Note that the pair $(k,j)$ is always in
$\mtA$; hence, $K_{j,m}^k\geq 1$ for every $j$, $k$ and $m$. Assume,
without loss of generality, that the pair $(k,j)$ is the first
element of the set $\mtA$.

Let $q(i)$ and $a(i)$ represent, respectively, the first and the
second components of the $i$th pair in set $\mtA$ for
$i=1,\ldots,K_{j,m}^k$. Then, (\ref{e. pulse model - complicated})
can be further simplified as follows:
   \begin{gather}
      r_{j,m}^k = h_{l_m^k}^kb_j^k[\Sv_k]_{jN_c+c_j^k,\lfloor
      j/N_f \rfloor} + \tilde{\hv}_{j,m}^k\tilde{\bv}_{j,m}^k +
      n_{l(j,k,m)},
      \label{e. pulse model - simple}
   \end{gather}
where $\tilde{\hv}_{j,m}^k = \Bigg{[}
      \left[\Sv_{q(2)}\right]_{a(2)N_c+c_{a(2)}^{q(2)} ,\lfloor a(2)/N_f
      \rfloor} h_{l(j,k,m)-a(2)N_c-c_{a(2)}^{q(2)}}^{q(2)} , \\
      \ldots\left. , \left[\Sv_{q\(K_{j,m}^k\)}\right]_{a\(K_{j,m}^k\)N_c+c_{a\(K_{j,m}^k\)}^{q\(K_{j,m}^k\)}
      ,\lfloor a\(K_{j,m}^k\)/N_f \rfloor}
      h_{l(j,k,m)-a\(K_{j,m}^k\)N_c-c_{a\(K_{j,m}^k\)}^{q\(K_{j,m}^k\)}}^{q\(K_{j,m}^k\)}
      \right]$ and $\tilde{\bv}_{j,m}^k = \left[b_{a(2)}^{q(2)}, \ldots,
      b_{a(K_{j,m}^K)}^{q(K_{j,m}^k)} \right]^T$.

From (\ref{e. combSample}) and (\ref{e. pulse model - simple}),
$\tilde{r}_j^k$ can be expressed as
\begin{gather}\label{e. rTilde_simple}
\tilde{r}_j^k=A\,b_j^k+\sum_{m=1}^{M}
h_{l_m^k}^k\tilde{\hv}_{j,m}^k\tilde{\bv}_{j,m}^k+\tilde{n}_j^k,
\end{gather}
where $A=[\Sv_k]_{jN_c+c_j^k,\lfloor
j/N_f\rfloor}\sum_{m=1}^{M}\left(h_{l_m^k}^k\right)^2$, and
$\tilde{n}_j^k=\sum_{m=1}^{M}h_{l_m^k}^kn_{l(j,k,m)}$, which is
distributed as ${\mathcal{N}}\left(0\,,\,\tilde{\sigma}^2\right)$
with
$\tilde{\sigma}^2=\sigma_n^2\sum_{m=1}^{M}\left(h_{l_m^k}^k\right)^2$.

Based on (\ref{e. rTilde_simple}), the log-likelihood of
$\tilde{r}_j^k$ given $b_j^k$ is,
   \begin{gather}
      \log f\( \tilde{r}_j^k | b_j^k \) = C +
      \log\sum_{\check{\bv}\in \{\pm 1\}^{\tilde{K}_j^k}}
      \exp\left\{-\frac{1}{2\tilde{\sigma}^2}\(\tilde{r}_j^k
      -A\,b_j^k
      -\sum_{m=1}^{M}h_{l_m^k}^k\tilde{\hv}_{j,m}^k\tilde{\bv}_{j,m}\)^2\right\}\Pr(\check{\bv}),
      \label{e. apriori bit1}
   \end{gather}
where $C$ is a constant independent of $j$ and $k$, $\check{\bv}$ is
a vector comprised of the distinct $b_n^l$'s in
$\tilde{\bv}_{j,1}^k,\ldots,\tilde{\bv}_{j,M}^k$, and
$\tilde{K}_j^k$ is the size of $\check{\bv}$. Note that
$\tilde{K}_j^k$ represents the total number of pulses that have at
least one multipath component arriving at the receiver at the same
time as one of the sampled signal paths originating from the $j$th
pulse of the $k$th user. Also note that for a given value of
$\check{\bv}$, $\tilde{\bv}_{j,m}^k$ in (\ref{e. apriori bit1}) is
uniquely defined, and $\Pr(\check{\bv})$ is the {\em a priori}
probability, which is obtained from the extrinsic information
provided by the symbol detector. Since the extrinsic information
from the symbol detector is the following LLR, $\lambda_2^{n-1} \(
b_i^l \) =  \log \frac{\Pr \( b_i^l = 1 \) }{ \Pr \( b_i^l = -1 \)
}$ [cf. (\ref{eq:LL_symDet})], it can be shown, with the aid of some
algebraic manipulations, that \cite{FishlerMUD:04}
   \begin{gather}
      \Pr(\check{\bv}) = \frac{1}{2^{\tilde{K}_j^k}}\prod_{i=1}^{\tilde{K}_j^k}
      \left[ 1 + [\check{\bv}]_i\tanh \(
      \frac{1}{2}\lambda_2^{n-1}\left([\check{\bv}]_i\right)\) \right].
      \label{e. apriori bit2}
   \end{gather}

From (\ref{e. apriori bit1}) and (\ref{e. apriori bit2}), the {\em a
priori} LLR of $b_j^k$ can be written as follows:
   \begin{align}\nonumber
      & \log \frac{f\(\tilde{r}_j^k | b_j^k =1 \) }
         {f\(\tilde{r}_j^k | b_j^k = -1 \)}
         \defi \lambda_1^n\( b_j^k \) \\
      & = \log \frac{\sum_{\check{\bv} \in \{\pm 1\}^{\tilde{K}_j^k}}
      e^{-\frac{1}{2\tilde{\sigma}^2}\(\tilde{r}_j^k
      - A
      - \sum_{m=1}^{M}h_{l_m^k}^k\tilde{\hv}_{j,m}^k\tilde{\bv}_{j,m}^k\)^2}
      \prod_{i=1}^{\tilde{K}_j^k}
         \left[ 1 + [\check{\bv}]_i\tanh \(
            \frac{1}{2}\lambda_2^{n-1}\([\check{\bv}]_i\)\)\right]
      }
      {\sum_{\check{\bv} \in \{\pm 1\}^{\tilde{K}_j^k}}
      e^{-\frac{1}{2\tilde{\sigma}^2}\(\tilde{r}_j^k
      + A
      - \sum_{m=1}^{M}h_{l_m^k}^k\tilde{\hv}_{j,m}^k\tilde{\bv}_{j,m}^k\)^2}
      \prod_{i=1}^{\tilde{K}_j^k}
         \left[ 1 + [\check{\bv}]_i\tanh \(
            \frac{1}{2}\lambda_2^{n-1}\([\check{\bv}]_i\)\)
            \right]}.
      \label{e. apriori bit4}
   \end{align}
From (\ref{e. apriori bit4}) and (\ref{e. P-D L_1}), it is observed
that the {\em a posteriori} LLR is given by the sum of the prior
information obtained from the symbol detector and the extrinsic
information.

\subsection{The Symbol Detector}

The symbol detector exploits the fact that $b_{(i-1)N_f+1}^k =
\cdots = b_{iN_f}^k$ for every $k=1,\ldots,K$ and $i=1,\ldots, P$.
Therefore, the symbol detector computes the {\em{a posteriori}} LLR
of $b_j^k$ given the extrinsic information from the pulse detector,
and given $b_{(i-1)N_f+1}^k = \cdots = b_{iN_f}^k$ for every
$k=1,\ldots,K$ and $i=1,\ldots, P$. It can be shown that this LLR
has the following general structure \cite{FishlerMUD:04}:
   \begin{gather}\label{eq:LL_symDet}
      L_2^n(b_j^k) \defi \log \frac {\Pr \( b_j^k = 1 |
      \{ \lambda_1^n(b_j^k)\}_{j=1,k=1}^{PN_f,K};\mbox{constraints on pulses} \)}
      {\Pr \( b_j^k = -1 |
      \{ \lambda_1^n(b_j^k)\}_{j=1,k=1}^{PN_f,K};\mbox{constraints on pulses} \)}
      = \underbrace{\sum_{i=N_f\lfloor (j-1) / N_f \rfloor +1,i\neq j}^{N_f\lfloor (j-1) / N_f \rfloor + N_f}
       \lambda_1^{n}(b_i^k)}_{\lambda_2^{n}(b_j^k)} +
       \lambda_1^n(b_j^k),
   \end{gather}
where the constraints are $b_{(i-1)N_f+1}^k = \cdots = b_{iN_f}^k$
for every $k=1,\ldots,K$ and $i=1,\ldots, P$. In
(\ref{eq:LL_symDet}), the {\em a posteriori} LLR at the output of
the symbol detector is expressed as the sum of the prior information
from the pulse detector, $\lambda_1^n(b_j^k)$, and the extrinsic
information about $b_j^k$, denoted by $\lambda_2^n(b_j^k)$. This
extrinsic information is obtained from the information about all the
pulses except the $j$th pulse of the $k$th user. In the next
iteration this information is fed back to the pulse detector as {\em
a priori} information about the $j$th pulse of the $k$th user.

Note that the structure of the pulse-symbol detector is similar to
the joint-over-antenna turbo receiver in \cite{Vesel:06}, which
employs multiple turbo loops for each antenna, by considering
``composite'' modulation for multiple antennas as the inner code,
and channel coding for different users as the outer code. The main
differences are that, for the pulse-symbol detector, the outer code
is a simple repetition code, while the inner code is a binary phase
shift keying modulation, and that there are also TH and polarity
randomization operations in the pulse-symbol detector, which randomize the positions and the polarities of the pulses in different frames.

\subsection{Complexity}

It is easily seen that computing $\lambda_1\( b_j^k \)$ of (\ref{e.
apriori bit4}) is the most complex task in the pulse-symbol
detector. The complexity of computing $\lambda_1\(b_j^k\)$ is
exponential in the total number $\tilde{K}_j^k$ of pulses that have
at least one multipath component arriving at the receiver at the
same time as one of the sampled signal paths originating from the
$j$th pulse of the $k$th user. That is, as can be observed from
(\ref{e. apriori bit4}), the complexity of computing $\lambda_1\(
b_j^k \)$ is ${\mathcal O}\( 2^{\tilde{K}_j^k} \)$. Since there are
$N_f$ pulses per symbol per user, the complexity of one iteration
per symbol per user is easily seen to be ${\mathcal
O}\(\sum_{j=1}^{N_f} 2^{\tilde{K}_j^k}\) = {\mathcal
O}\(2^{Y(K)}\)$, where $Y(K) \defi \max_{j=1, \ldots,
N_f}\tilde{K}_j^k$. Denoting by $N_{\rm{i}}$ the number of
iterations made by the pulse-symbol detector, the complexity of the
pulse-symbol detector is ${\mathcal O}\(N_{\rm{i}}2^{Y(K)}\)$ per
symbol per user.

$\tilde{K}_j^k$ is a random variable depending on the channel
impulse response, the TH sequence, and the number of users in the
system. It is hard to compare the complexity of the pulse-symbol
detector, which is random, with the complexity of multiuser
detection algorithms that have fixed complexity, e.g., the optimal
detector. Nevertheless, if, for example, the probability of the
event $N_{\rm{i}}2^{Y(K)}>2^K$ is very low, then, roughly speaking,
the proposed algorithm is simpler than the optimal detector.

The exact distribution of $Y(K)$ is very complicated, and moreover,
this distribution depends on the exact channel structure, the number
of paths arriving at the receiver, and the TH sequences. In what
follows, numerical examples are used to demonstrate the complexity
of the pulse-symbol detector. In particular, consider a system with
$20$ users, each transmitting at rate of $2$ MBits/sec over a $0.5$
GHz UWB indoor channel \cite{IEEE802153a_chan}. The receiver is
sampling the first $10$ multipath components; i.e.,
$\Lcal=\{1,2,\ldots,10\}$. Figure \ref{fig2} depicts the empirical
cumulative distribution function (CDF) of $Y(K)$, averaged over
$100$ different channel realizations from the channel model $1$
(CM-1) of the IEEE 802.15.3a channel model, for systems transmitting
one, five and twenty pulses per symbols ($N_f=1,5,20$). It is clear
that the complexity of the pulse-symbol detector decreases as the
pulse rate, $N_f$, decreases. This is expected because, as the pulse
rate decreases, the probability of collisions decreases as well,
which reduces the complexity of the pulse-symbol detector.
Nevertheless, the complexity of the pulse-symbol detector can be
large even for moderate numbers of pulses per symbol. In the next
section, two low-complexity implementations are presented.

\section{Low Complexity Implementations}

The complexity of the pulse-symbol detector varies considerably with
the system pulse rate, $N_f$. An increase in the pulse rate
increases the algorithm complexity, and this complexity can be large
even for moderate pulse rates or numbers of users. In what follows
two low complexity implementations are described. The first one is
based on approximating part of the multiple access interference
(MAI) by a Gaussian random variable, while the second one is based
on soft interference cancellation.

\subsection{Low-Complexity Implementation: The Gaussian
Approximation Approach}\label{sec:LC}

The high complexity of the pulse-symbol detector is due solely to
the pulse detector where the {\em a priori} LLR of a received sample
given the transmitted symbol, $\lambda_1(b_j^k)$, is computed. In
recent studies (see,
\cite{Cassioli:02,Turin:02,Win:02,IEEE802153a_chan}, and references
therein), UWB channels are commonly characterized as multipath
channels with large numbers of paths, and delay spreads of up to a
few tens of nanoseconds. These large delay spreads are equivalent to
discrete-time channels having more than one hundred taps. Although
the UWB channel consists of many taps, most of them are weak
compared with the strongest tap, and only about five to ten taps are
weaker by no more than $10$ dB than the strongest tap. Therefore,
most of the pulses colliding with the pulse of interest arrive via
weak paths.

In order to reduce the complexity of the pulse-symbol detector, we
propose to model the MAI resulting from the pulses arriving via weak
paths by a Gaussian random variable. Recall that $h_{l_m^k}^k$ is
the gain of the $m$th path, through which the pulse of interest
arrives at the receiver. In order to reduce the complexity of
computing $\lambda_1^n\(b_j^k\)$, the receiver sets a threshold $T$
(in dB) and all the pulses colliding with the pulse of interest are
divided into two groups. The first group contains all the pulses
that collide with the pulse of interest and that arrive via paths
that are weaker than the $m$th path of user $k$ by no more than $T$
dB (i.e., each path has an amplitude of at least
$10\log_{10}\left|h_{l_m^k}^k\right|-T$ dB). The second group
contains all the pulses that collide with the pulse of interest and
that arrive via paths that are weaker than $h_{l_m^k}^k$ by more
than $T$ dB. Denote by $I_{j,m}^k$ and $\bar{I}_{j,m}^k$ the indices
of the pulses belonging to the first and second group, respectively;
that is,
   \begin{gather}
      I_{j,m}^k = \left\{i \,
      \Big{|} \,
      10\log_{10}\left| h_{l_m^k}^k \right|
      - 10\log_{10}\left|h_{l(j,k,m)-a(i)N_c-c_{a(i)}^{q(i)}}^{q(i)}\right|
      \le T,i=2,\ldots,K_{j,m}^k \right\},
   \end{gather}
and similarly define $\bar{I}_{j,m}^k$.

A model for $r_{j,m}^k$ can be written in terms of $I_{j,m}^k$ and
$\bar{I}_{j,m}^k$ as follows:
   \begin{align}
      r_{j,m}^k &= h^k_{l_m^k}b_j^k[\Sv_k]_{jN_c+c_j^k,\lfloor
      j/N_f \rfloor} + \sum_{i\in I_{j,m}^k}
      b_{a(i)}^{q(i)}
      \left[\Sv_{q(i)}\right]_{a(i)N_c+c_{a(i)}^{q(i)}
      ,\lfloor a(i)/N_f \rfloor}
      h_{l(j,k,m)-a(i)N_c-c_{a(i)}^{q(i)}}^{q(i)} \nonumber \\
      &+
      \sum_{i\in \bar{I}_{j,m}^k}
      b_{a(i)}^{q(i)}
      \left[\Sv_{q(i)}\right]_{a(i)N_c+c_{a(i)}^{q(i)}
      ,\lfloor a(i)/N_f \rfloor}
      h_{l(j,k,m)-a(i)N_c-c_{a(i)}^{q(i)}}^{q(i)}
      + n_{l(j,k,m)},
      \label{e. pulse model - simple2}
   \end{align}
where the first term on the RHS represents the part of the received
signal resulting from the pulse of interest, the second term on the
RHS represents that part of the MAI resulting from strong
interference, the third term on the RHS represents that part of the
MAI resulting from weak interference, and the fourth term on the RHS
represents the additive Gaussian noise. Since most of the paths are
considerably weaker than the main path, it is expected that
$|\bar{I}_{j,m}^k| >> |I_{j,m}^k|$. As such, the third term on the
RHS of (\ref{e. pulse model - simple2}) is the sum of a large number
of random variables and we propose to model this sum as a Gaussian
random variable. The mean and the variance of the third term on the
RHS of (\ref{e. pulse model - simple2}) are zero and $\sum_{i\in
\bar{I}_{j,m}^k}
\left|h_{l(j,k,m)-a(i)N_c-c_{a(i)}^{q(i)}}^{q(i)}\right|^2$,
respectively. Thus we use the following approximation:
   \begin{eqnarray}
      \sum_{i\in \bar{I}_{j,m}^k}
      b_{a(i)}^{q(i)}
      \left[\Sv_{q(i)}\right]_{a(i)N_c+c_{a(i)}^{q(i)}
      ,\lfloor a(i)/N_f \rfloor}
      h_{l(j,k,m)-a(i)N_c-c_{a(i)}^{q(i)}}^{q(i)} \sim \N{0,
      \sum_{i\in \bar{I}_{j,m}^k}
      \left|h_{l(j,k,m)-a(i)N_c-c_{a(i)}^{q(i)}}^{q(i)}\right|^2}.
      \label{e. approximation weak MAI}
   \end{eqnarray}

Approximating the part of the MAI corresponding to weak pulses
colliding with the pulse of interest by a Gaussian random variable
results in the following approximate model for $r_{j,m}^k$:
   \begin{align}
      r_{j,m}^k &\approx h^k_{l_m^k}b_j^k[\Sv_k]_{jN_c+c_j^k,\lfloor
      j/N_f \rfloor} + \sum_{i\in I_{j,m}^k}
      b_{a(i)}^{q(i)}
      \left[\Sv_{q(i)}\right]_{a(i)N_c+c_{a(i)}^{q(i)}
      ,\lfloor a(i)/N_f \rfloor}
      h_{l(j,k,m)-a(i)N_c-c_{a(i)}^{q(i)}}^{q(i)}
      + \check{n}_{j,m}^k  \nonumber \\
      &=
      h^k_{l_m^k}b_j^k[\Sv_k]_{jN_c+c_j^k,\lfloor
      j/N_f \rfloor} +
      \dtilde{\hv}_{j,m}^k\dtilde{\bv}_{j,m}^k
      + \check{n}_{j,m}^k,
      \label{e. pulse model - approx}
   \end{align}
where $\check{n}_{j,m}^k$ is a zero mean Gaussian random variable
with variance $(\sigma_{j,m}^{k})^2= \sigman + \sum_{i\in
\bar{I}_{j,m}^k}
\left|h_{l(j,k,m)-a(i)N_c-c_{a(i)}^{q(i)}}^{q(i)}\right|^2$;
$\dtilde{\hv}_{j,m}^k = \left[
\left[\Sv_{q(I_1)}\right]_{a(I_1)N_c+c_{a(I_1)}^{q(I_1)} ,\lfloor
a(I_1)/N_f \rfloor}
h_{l(j,k,m)-a(I_1)N_c-c_{a(I_1)}^{q(I_1)}}^{q(I_1)} ,\right. \\
\ldots\left.
,\left[\Sv_{q\(I_{|I|}\)}\right]_{a\(I_{|I|}\)N_c+c_{a\(I_{|I|}\)}^{q\(I_{|I|}\)}
,\lfloor a\(I_{|I|}\)/N_f \rfloor}
h_{l(j,k,m)-a\(I_{|I|}\)N_c-c_{a\(I_{|I|}\)}^{q\(I_{|I|}\)}}^{q\(I_{|I|}\)}
\right]$ and $\dtilde{\bv}_{j,m}^k = \left[b_{a(I_1)}^{q(I_1)},
\ldots, b_{a(I_{|I|})}^{q(I_{|I|})} \right]$. Using the same
derivations leading to (\ref{e. apriori bit4}) and (\ref{e. pulse
model - approx}), the {\em a priori} log-likelihood ratio of
$\tilde{r}_j^k=\sum_{m=1}^{M}h_{l_m^k}^kr_{j,m}^k$ given $b_j^k$ is
then approximated by,
   \begin{eqnarray}
      &&\tilde{\lambda}_1^n\(b_j^k \)
      =\log \frac{f\(\tilde{r}_j^k | b_j^k =1 \) }
         {f\(\tilde{r}_j^k | b_j^k = -1 \)} \cong\\\nonumber
      &&\log \frac{\sum_{\check{\check{\bv}} \in \{\pm 1\}^{\tilde{\tilde{K}}_j^k}}
      e^{-\frac{1}{2\tilde{\tilde{\sigma}}^2}\(\tilde{r}_j^k
      -\tilde{A}
      - \sum_{m=1}^{M}h_{l_m^k}^k\tilde{\tilde{\hv}}_{j,m}^k\tilde{\tilde{\bv}}_{j,m}^k\)^2}
      \prod_{i=1}^{\tilde{\tilde{K}}_j^k}
         \left[1+[\check{\check{\bv}}]_i\tanh \(
            \frac{1}{2}\lambda_2^{n-1}\([\check{\check{\bv}}]_i\)\)
            \right] }
         {\sum_{\check{\check{\bv}}\in \{\pm 1\}^{\tilde{\tilde{K}}_j^k}}
      e^{-\frac{1}{2\tilde{\tilde{\sigma}}^2}\(\tilde{r}_j^k
      +\tilde{A}
      - \sum_{m=1}^{M}h_{l_m^k}^k\tilde{\tilde{\hv}}_{j,m}^k\tilde{\tilde{\bv}}_{j,m}^k\)^2}
      \prod_{i=1}^{\tilde{\tilde{K}}_j^k}
         \left[1+[\check{\check{\bv}}]_i\tanh\(
            \frac{1}{2}\lambda_2^{n-1}\([\check{\check{\bv}}]_i\)\)
            \right] },
      \label{e. apriori bit5}
   \end{eqnarray}
where $\tilde{A}=[\Sv_k]_{jN_c+c_j^k,\lfloor j/N_f
\rfloor}\sum_{m=1}^{M}\left(h_{l_m^k}^k\right)^2$,
$\tilde{\tilde{\sigma}}^2$ is the variance of
$\sum_{m=1}^{M}h_{l_m^k}^k\check{n}_{j,m}^k$, which is
$\sum_{m=1}^{M}|h_{l_m^k}^k|^2(\sigma_{j,m}^{k})^2$,
$\check{\check{\bv}}$ is a vector comprised of the distinct $b_n^l$'s in $\tilde{\tilde{\bv}}_{j,1}^k,\ldots,\tilde{\tilde{\bv}}_{j,M}^k$,
and $\tilde{\tilde{K}}_j^k$ is the size of $\check{\check{\bv}}$.

The proposed low complexity implementation computes the approximate
{\em a priori} log-likelihood ratios, $\left\{\tilde{\lambda}_1^n\(
b_j^k \)\right\}$, instead of the exact {\em a priori}
log-likelihood ratios. The symbol detector uses these approximate
LLRs as the extrinsic information, and it computes a new set of
extrinsic information variables, $\{ \lambda_2^n(b_j^k)\}$, based on
the approximate LLRs provided by the pulse detector. The algorithm
continues to iterate between the two stages until convergence is
reached.

The complexity of the proposed scheme depends on the exact number of
strong pulses colliding with the pulse of interest, which is again a
random variable. It is easily seen that the complexity of this
implementation is ${\mathcal O}\( 2^{\tilde{Y}(K)} \)$, where
$\tilde{Y}(K)=\max_{j=1, \ldots, N_f }\tilde{\tilde{K}}_j^k$. Again,
we resort to a numerical example in order to demonstrate the
complexity of the proposed detector. Consider a system having $20$
users, each transmitting at a rate of $2$ MBits/sec over a $0.5$ GHz
UWB indoor channel \cite{IEEE802153a_chan}. The receiver is sampling
the first $10$ multipath components; i.e.,
$\Lcal=\{1,2,\ldots,10\}$, and the threshold $T$ is set to $3$ dB.
Figure \ref{fig3} depicts the empirical CDF of $\tilde{Y}(K)$,
averaged over $100$ different channel realizations from the channel
model $1$ (CM-1) of the IEEE 802.15.3a channel model, for systems
transmitting one, five and twenty pulses per symbols ($N_f=1,5,20$).
By comparing Figure \ref{fig2} and Figure \ref{fig3}, the reduction
in the complexity compared with the complexity of the pulse-symbol
detector can be observed. In Figure \ref{fig3_2}, the empirical CDF
is plotted for $N_f=5$ and various threshold values. It is observed
that as the threshold is decreased, fewer collisions are considered
as strong ones, which reduces the complexity of the algorithm.

Using the same approach, there are other ways of reducing the
complexity of the pulse-symbol detector. For example, one can
divide the received pulses into two groups based on their relative
strengths. In this approach, a threshold $\delta$ will be set in
advance, and the MAI caused by all but the $\delta$ strongest
colliding pulses will be modelled as a Gaussian random variable.
In this approach the complexity of the receiver is limited by
$N_f2^{\delta}$ per symbol per user.

\subsection{Low-Complexity Implementation:
The Soft Interference Cancellation Approach}

The complexity of the low-complexity implementation presented in
the previous subsection might still be high for large numbers of
users or pulse rates. As such, an even simpler implementation
method is required. In what follows a very low complexity
implementation based on soft interference cancellation is
presented.

Recall that the most complex task in the pulse-symbol detector is
the computation of the {\em a priori} log-likelihood ratio of the
received sample given the transmitted pulse, $\lambda_1\( b_j^k\) =
\log \frac{f\( \tilde{r}_j^k | b_j^k =1 \) }{f\( \tilde{r}_j^k |
b_j^k = -1 \) }$. Our aim is to find a simple way to approximate
$\lambda_1\( b_j^k \)$, and soft-interference cancellation provides
us with such a method \cite{Hagenauer:96b,Lampe:99}. Recall that the
model for $\tilde{r}_j^k$ is given by
$\tilde{r}_j^k=\sum_{m=1}^{M}r_{j,m}^k$, where $ r_{j,m}^k =
h_{l_m^k}^kb_j^k[\Sv_k]_{jN_c+c_j^k,\lfloor j/N_f \rfloor} +
\tilde{\hv}_{j,m}^k\tilde{\bv}_{j,m}^k + n_{l(j,k,m)}$. In
soft-interference cancellation methods, the first step is to form a
soft estimate of $\tilde{\bv}_{j,m}^k$. This soft estimate is the
conditional mean of $\tilde{\bv}_{j,m}^k$ based on our current
knowledge. We denote this soft estimate by $\bartilde{\bv}_{j,m}^k =
{\rm{E}}\left\{\tilde{\bv}_{j,m}^k \big{|}
\{\lambda_2\(b_j^k\)\}\right\}$, which is given by
   \begin{align}
      &\left[\bartilde{\bv}_{j,m}^k \right]_i = \left[
      {\rm{E}}\left\{\tilde{\bv}_{j,m}^k | \{\lambda_2\(b_j^k\)\}\right\}\right]_i
      = {\rm{E}}\left\{b_{a(i)}^{q(i)}\right\}
      = \Pr\( b_{a(i)}^{q(i)} = 1 \) - \Pr\( b_{a(i)}^{q(i)} = -1 \)
      \nonumber \\
      & \quad\quad = \frac{1}{2}\left[1 + \tanh\( \frac{1}{2}\lambda_2\(
      b_{a(i)}^{q(i)} \) \) \right] - \frac{-1}{2}\left[1 - \tanh\( \frac{1}{2}\lambda_2\(
      b_{a(i)}^{q(i)} \) \) \right] = \tanh\( \frac{1}{2}\lambda_2\(
      b_{a(i)}^{q(i)} \) \).
   \end{align}
Assuming that this soft estimate is reliable, the remodulated
signal $\tilde{\hv}_{j,m}^k\bartilde{\bv}_{j,m}^k$ is subtracted
from $r_{j,m}^k$ resulting in
   \begin{eqnarray}\label{e. r_softEst}
      \bar{r}_{j,m}^k \defi r_{j,m}^k -
      \tilde{\hv}_{j,m}^k\bartilde{\bv}_{j,m}^k = h_{l_m^k}^kb_j^k[\Sv_k]_{jN_c+c_j^k,\lfloor j/N_f
      \rfloor} + \tilde{\hv}_{j,m}^k\(\tilde{\bv}_{j,m}^k - \bartilde{\bv}_{j,m}^k \) +
      n_{l(j,k,m)}.
   \end{eqnarray}
Subtracting the remodulated signal from $r_{j,m}^k$ results in the
reduction of the MAI. Since the number of collisions is large, the
remaining MAI, $\tilde{\hv}_{j,m}^k\(\tilde{\bv}_{j,m}^k -
\bartilde{\bv}_{j,m}^k \) =
\sum_{i=2}^{K_{j,m}^k}\left[\tilde{\hv}_{j,m}^k\right]_i \(
b_{a(i)}^{q(i)} - {\rm{E}}\left\{b_{a(i)}^{q(i)}\right\} \)$, is
approximated by a Gaussian random variable, as follows:
   \begin{gather}
      \sum_{i=2}^{K_{j,m}^k}\left[\tilde{\hv}_{j,m}^k\right]_i \(
      b_{a(i)}^{q(i)} - {\rm{E}}\left\{b_{a(i)}^{q(i)}\right\} \) \sim
      \N{\mu_{j,m}^k,(\sigma_{j,m}^{k})^2 }
      \label{e. dist SIC}
   \end{gather}
with
   \begin{gather}
      \mu_{j,m}^k = {\rm{E}}\left\{ \sum_{i=2}^{K_{j,m}^k}\left[\tilde{\hv}_{j,m}^k\right]_i \(
      b_{a(i)}^{q(i)} - \E{ b_{a(i)}^{q(i)} } \tilde{\bv}_{j,m}^k \) \right\}
      = \sum_{i=2}^{K_{j,m}^k}\left[\tilde{\hv}_{j,m}^k\right]_i {\rm{E}}\left\{
      \( b_{a(i)}^{q(i)} - \E{ b_{a(i)}^{q(i)} } \) \right\} = 0
      \label{e. mean SIC}
   \end{gather}
and
   \begin{align}
      (\sigma_{j,m}^{k})^2&={\rm{Var}}\left\{ \sum_{i=2}^{K_{j,m}^k}\left[\tilde{\hv}_{j,m}^k\right]_i \(
      b_{a(i)}^{q(i)} - {\rm{E}}\left\{b_{a(i)}^{q(i)}\right\}  \) \right\}
      = {\rm{E}}\left\{\( \sum_{i=2}^{K_{j,m}^k}\left[\tilde{\hv}_{j,m}^k\right]_i \(
      b_{a(i)}^{q(i)} - {\rm{E}}\left\{b_{a(i)}^{q(i)}\right\} \) \)^2 \right\} \nonumber\\
      &=\sum_{i=2}^{K_{j,m}^k}\left[\tilde{\hv}_{j,m}^k\right]_i{\rm{Var}}\left\{b_{a(i)}^{q(i)}\right\}
      = \sum_{i=2}^{K_{j,m}^k}\left[\tilde{\hv}_{j,m}^k\right]^2_i\( 1 -
      \( \left[ \dtilde{b}_j^k \right]_i \)^2 \),
      \label{e. var SIC}
   \end{align}
where ${\rm{E}}\left\{\( b_{a(i)}^{q(i)} -
{\rm{E}}\left\{b_{a(i)}^{q(i)}\right\} \)\( b_{a(l)}^{q(l)} -
{\rm{E}}\left\{b_{a(l)}^{q(l)}\right\} \)\right\} = 0$ for $i \neq
l$, and ${\rm{Var}}\left\{b_{a(i)}^{q(i)}\right\} =
{\rm{E}}\left\{\(b_{a(i)}^{q(i)}\)^2\right\} -
\({\rm{E}}\left\{\(b_{a(i)}^{q(i)}\)\right\} \)^2 = 1 - \( \left[
\dtilde{b}_j^k \right]_i \)^2$ are used.

Then, the soft estimate for $\tilde{r}_j^k$ can be obtained as
\begin{gather}
\bar{\tilde{r}}_j^k=\sum_{m=1}^{M}h_{l_m^k}^k\bar{r}_{j,m}^k
=\tilde{A}\,b_j^k+\bar{\tilde{n}}_j^k,
\end{gather}
where $\tilde{A}=[\Sv_k]_{jN_c+c_j^k,\lfloor
j/N_f\rfloor}\sum_{m=1}^{M}\left(h_{l_m^k}^k\right)^2$, and
$\bar{\tilde{n}}_j^k=\sum_{m=1}^{M}h_{l_m^k}^k\bar{n}_{j,m}^k$, with
$\bar{n}_{j,m}^k=\tilde{\hv}_{j,m}^k\(\tilde{\bv}_{j,m}^k -
\bartilde{\bv}_{j,m}^k \)+n_{l(j,k,m)}$.

In the proposed very low-complexity implementation of the
pulse-symbol algorithm, the pulse detector computes the {\em a
priori} log-likelihood ratio of $\bar{\tilde{r}}_j^k$ given the
transmitted symbol, instead of the {\em a priori} log-likelihood
ratio of $\tilde{r}_j^k$ given the transmitted symbol. Denote by
$\dtilde{\lambda}_1^n \( b_j^k \)$ this log-likelihood ratio; that
is, $ \dtilde{\lambda}_1^n\( b_j^k \)
\defi \log \frac{f\(\bar{\tilde{r}}_j^k | b_j^k = 1 \)}{ f\(
\bar{\tilde{r}}_j^k | b_j^k = -1 \)}$. By using the Gaussian
approximation for the residual MAI as shown in (\ref{e. dist SIC}),
$\dtilde{\lambda}_1^n\( b_j^k \)$ is easily seen to be given by
   \begin{eqnarray}
      \dtilde{\lambda}_1^n\( b_j^k \)=\frac{-\(\bar{\tilde{r}}_j^k -\tilde{A}\)^2
      + \( \bar{\tilde{r}}_j^k + \tilde{A} \)^2 }
      {\sum_{m=1}^{M}\left(h^k_{l_m^k}\right)^2
      \left(\sigman + (\sigma_{j,m}^{k})^2\right)}
      = \frac{4\tilde{A}\bar{\tilde{r}}_j^k}
      {\sum_{m=1}^{M}\left(h^k_{l_m^k}\right)^2
      \left(\sigman + (\sigma_{j,m}^{k})^2\right)}.
   \end{eqnarray}

As in the previously proposed low complexity implementation, the
pulse detector computes the {\em a priori} log-likelihood ratios,
$\left\{ \dtilde{\lambda}_1^n\( b_j^k \) \right\}$, instead of the
exact {\em a priori} log-likelihood ratios. The symbol detector
uses these approximated LLRs as its extrinsic information, and it
computes a new set of extrinsic information, $\{
\lambda_2^n(b_j^k)\}$, based on the approximated LLRs provided by
the pulse detector. The algorithm then continues to iterate
between the two stages until convergence is reached.

\section{Simulations}

In this section, simulation results are presented in order to
investigate the performance of various receiver structures as a
function of the signal-to-noise ratio (SNR). The UWB indoor channel
model reported by the IEEE 802.15.3a task group is used for
generating UWB multipath channels \cite{IEEE802153a_chan}, and the
uplink of a synchronous TH-IR system with $N_f=5$, $N_c=250$, and a bandwidth of $0.5$ GHz is considered. It is assumed that there is no inter-frame interference (IFI) in the system\footnote{TH codes are generated randomly from $\{0,1,\ldots,N_c-L-1\}$ in order not to cause any IFI.}. Note, however, that the analysis in Section III and IV cover scenarios with IFI, as well.

In Figure \ref{fig:BEP}, bit error rates (BERs) of various receivers
are plotted as functions of the SNR using $100$ realizations of
CM-1 \cite{IEEE802153a_chan}. There are $5$ users in the environment
($K=5$), where the first user is assumed to be the user of interest.
Each interfering user is modeled to have $10$ dB more power than the
user of interest so that an MAI-limited scenario can be
investigated. Note that the benefits of iterative multiuser
detectors become more obvious in the MAI-limited regime. At all the
receivers, the first $25$ multipath components are employed; i.e.,
${\Lcal}^1=\{1,\ldots,25\}$. In the figure, the curve labeled
``MRC-Rake'' corresponds to the performance of a conventional
MRC-Rake receiver \cite{Cassioli:2002}; the curves labeled ``LC''
correspond to the performance of the low complexity implementation
method based on the Gaussian approximation ($T=10$ dB is used); and
the curves labeled ``SIC'' correspond to the performance of the low
complexity implementation method based on soft interference
cancellation. Also, the single user bound is plotted for an MRC-Rake
receiver in the absence of interfering users. From the figure, it is
observed that the BERs of the proposed detectors are considerably
lower than those of the MRC-Rake. In addition, after two iterations, the performance of the proposed receivers gets very close to that of a single user system. Finally, the low complexity implementation based on the Gaussian
approximation out-performs the low complexity implementation based
on soft interference cancellation on the first iteration, which is a
price paid for the lower complexity of the latter algorithm. In
other words, the soft interference approach estimates the overall
MAI by first order moments, and approximates the difference between
the MAI and the MAI estimate by Gaussian random variables, which
reduces the complexity significantly but also causes a performance
loss due to a more extensive Gaussian approximation compared to the
low complexity implementation that uses Gaussian approximations only for weak MAI terms. However,
after two iterations, both receivers get very close to the
single-user bound, and the low complexity implementation based on
soft interference cancellation becomes more advantageous due to its
lower computation complexity (cf. Figure \ref{fig:BEP_comp}).

In Figure \ref{fig:BEP_thr}, the same parameters as in the previous
case are used, and performance of the low complexity implementation
based on the Gaussian approximation is investigated for various
threshold values. As can be observed from the plot, as the threshold
is decreased; i.e., as more MAI terms are approximated by Gaussian
random variables, the performance of the algorithm degrades. In
other words, there is a tradeoff between performance and complexity
as expected from the study in Section \ref{sec:LC}. Also note that
since each interfering user is $10$ dB stronger than the user of
interest, there is not much difference between the $T=10$ dB and
$T=0$ dB cases (as most of the significant MAI terms are usually
above the threshold in both cases), whereas the performance degrades
significantly for the $T=-10$ dB case.

Next, the performance of the receivers is investigated for CM-3 of
the IEEE 802.15.3a channel model, where $T=0$ dB is used for the low
complexity implementation based on the Gaussian
approximation\footnote{The curves are very similar to the ones in
Figure \ref{fig:BEP}; hence they are not shown separately.}. The same
observations as in Figure \ref{fig:BEP} are made. The main
difference in this case is the increase in the BERs, which is a
result of the larger channel delay spread of the channel model used
in the simulations. In other words, less energy is collected on the
average, which results in an increase in average BERs.

In order to compare the performance of the proposed receivers under
computational constraints, the performance loss (in dB) of each
receiver compared to a single user receiver is plotted versus the
average number of multiplication operations per user in Figure
\ref{fig:BEP_comp}. The performance loss is calculated as the
difference between the SNR needed for the receiver to achieve a BER
of $10^{-3}$ and the SNR of the single user receiver at
BER=$10^{-3}$. For each receiver, the points on the curve are
obtained for $1$, $2$ and $3$ iterations. From Figure
\ref{fig:BEP_comp}, it is concluded that the low complexity
implementation based on soft interference cancellation provides a
better performance-complexity tradeoff than the low complexity
implementation based on the Gaussian approximation.

Finally, the performance of the receivers that are sampling only the
first $5$ multipath components (i.e., ${\Lcal}^1=\{1,2,3,4,5\}$) is
investigated. In this case, it is observed from Figure
\ref{fig:BEP_5} that the proposed receivers can still perform very
closely to the single-user bound, whereas the MRC-Rake receiver
experiences a serious error floor.

\section{Summary and Concluding Remarks}

In this paper an iterative approach, the pulse-symbol detector, for
multiuser detection in TH-IR systems has been presented for
frequency-selective environments. In this approach, the detection
problem is divided, artificially, into two parts, and the proposed
algorithm iterates between these two parts. In each iteration, the
algorithm passes extrinsic information between the two parts,
resulting in an increase in the accuracy of the decisions made by
the detector. The complexity of the proposed detector is random;
hence, comparing the complexity of this detector with other fixed
complexity algorithms is complicated. Nevertheless, we have
demonstrated, via simulations, that there are scenarios were the
complexity of the proposed detector is lower than the complexity of
the optimal detector, while in others it is higher.

In addition, two low-complexity implementations have been presented.
The first implementation is based on approximating parts of the MAI
by a Gaussian random variable and the second is based on soft
interference cancellation. The complexity of both implementations is
quite low, and we believe that these algorithms could be used in
practical systems. The performance characteristics of these
low-complexity implementations have been examined using simulations.
We have shown that these algorithms typically get very close to the
single-user bound after only a few iterations, and outperform the
MRC-Rake substantially.

The proposed multiuser detection algorithms were described under the
assumption of synchronous users. However, it is easily seen that
this assumption was made only for notational simplicity. The pulse
detector inherently ignores any information about the symbols and
their structure, and in particular their timing. It uses only the
information about the individual pulses that collide with the pulse
of interest. The symbol detector uses the results of the pulse
detector for pulses that correspond to the symbol of interest. As
such, the symbol detector is independent of the other symbols from
the same user or from the symbols from other users. In summary, it
is evident that synchronization among users is not required.
Moreover, it is easy to design a serialized version of the proposed
algorithm in the sense that the receiver process on-the-fly new
samples at the expense of performance degradation. In summary, the
only requirement from the receiver is the knowledge of each user's
symbol timing, which is commonly obtained during synchronization
phases in practical systems.

\bibliographystyle{plain}
\bibliography{uwb,mud,rcdma}


\newpage

\begin{figure}
\begin{center}
\includegraphics[width=0.8\textwidth]{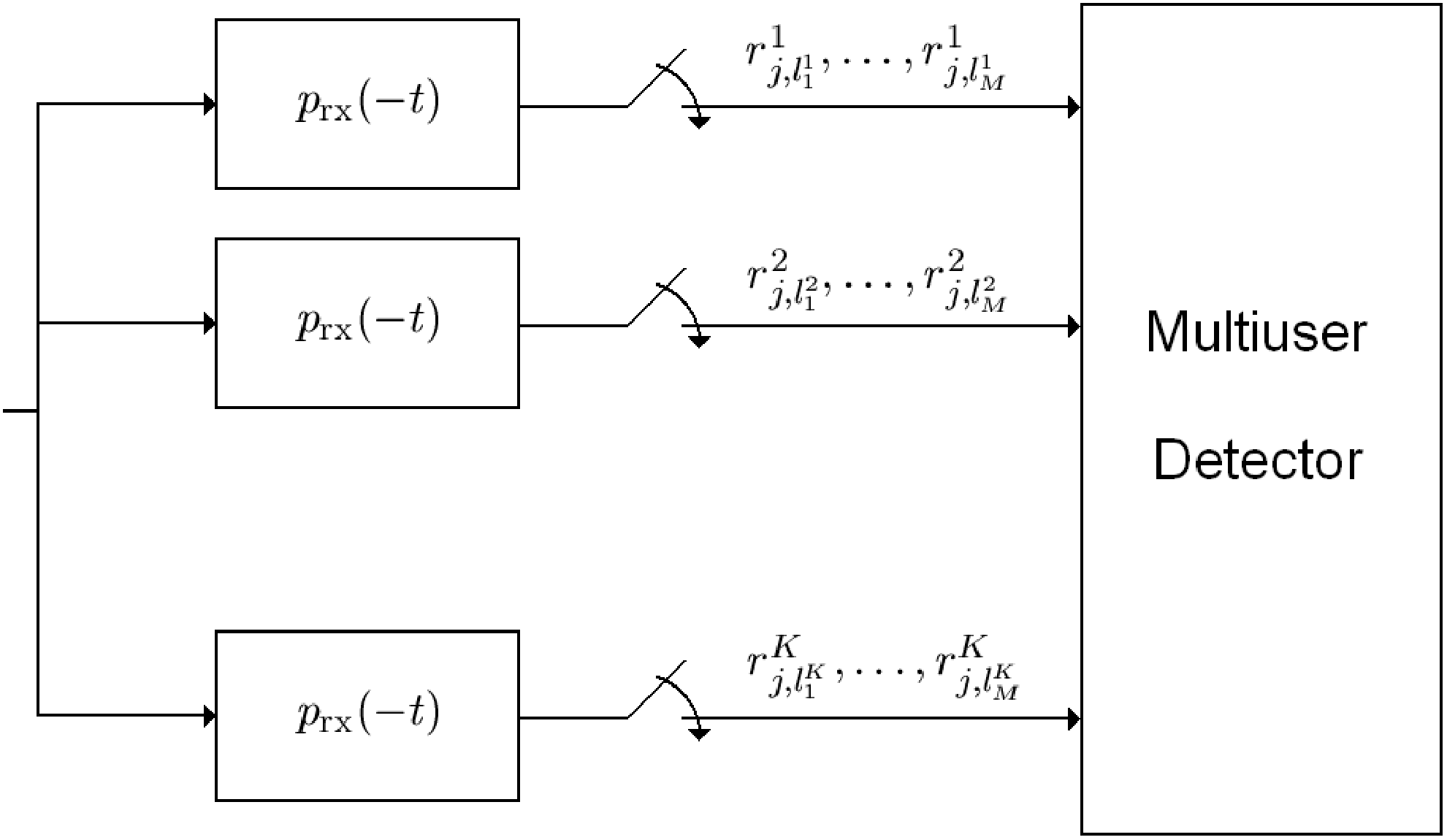}
\caption{The general structure of the receiver, where
$p_{\rm{rx}}(t)$ denotes the received UWB pulse.} \label{f: basic
receiver}
\end{center}
\end{figure}

\begin{figure}[ht]
   \includegraphics[scale=0.9]{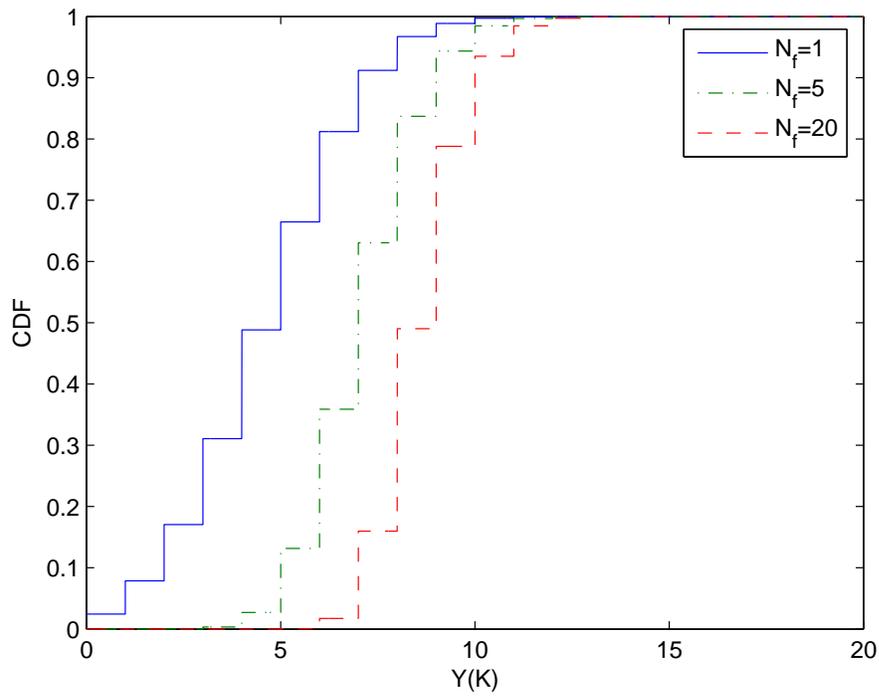}
   \caption{CDF of $\max_{j=1,\ldots,N_f} \tilde{K}_j^k$ for various pulse rates.}
   \label{fig2}
\end{figure}

\begin{figure}[ht]
   \includegraphics[scale=0.9]{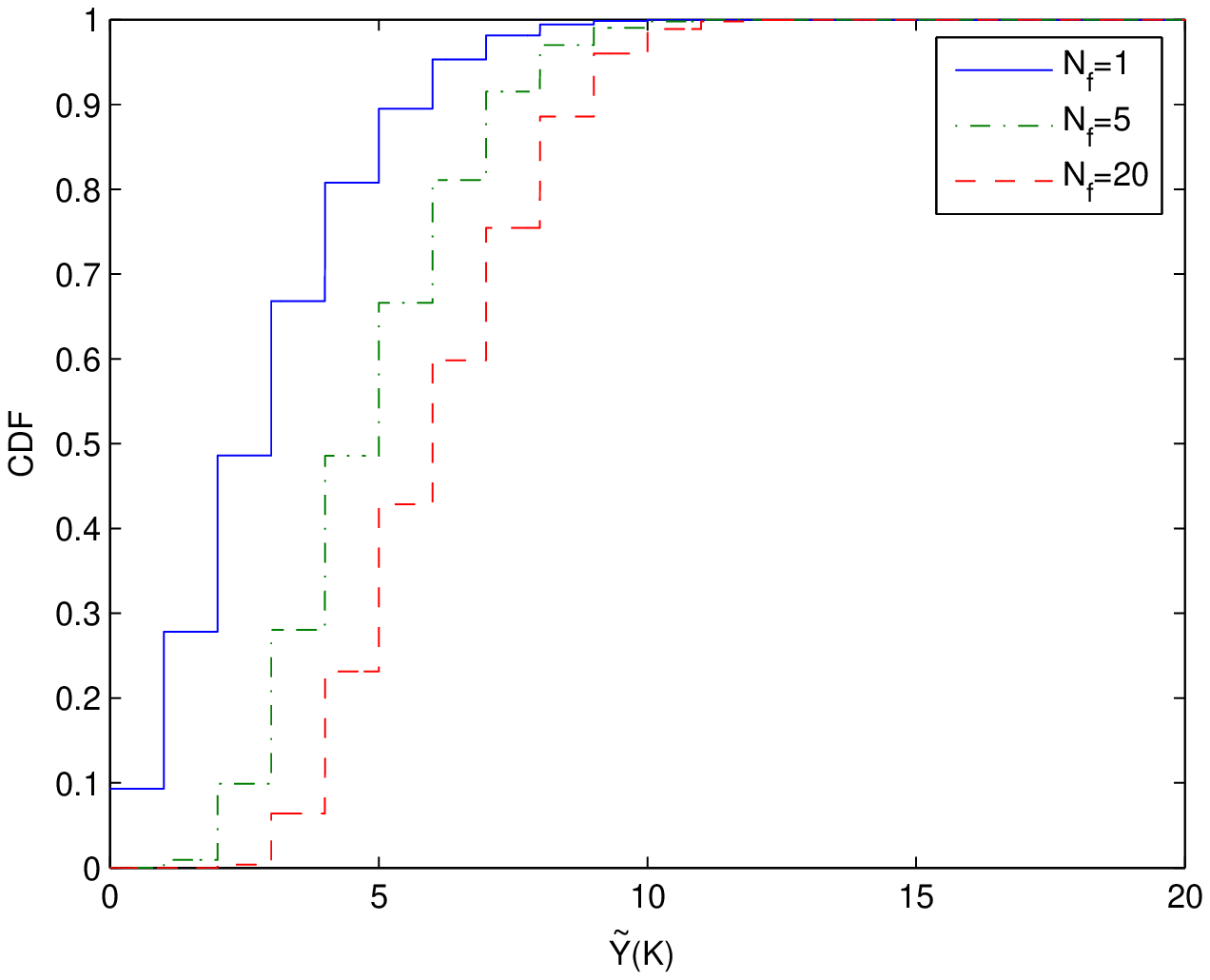}
   \caption{CDF of $\max_{j=1,\ldots, N_f} \tilde{\tilde{K}}_j^k$ for various pulse rates and $T=3$ dB.}
   \label{fig3}
\end{figure}

\begin{figure}[ht]
   \includegraphics[scale=0.9]{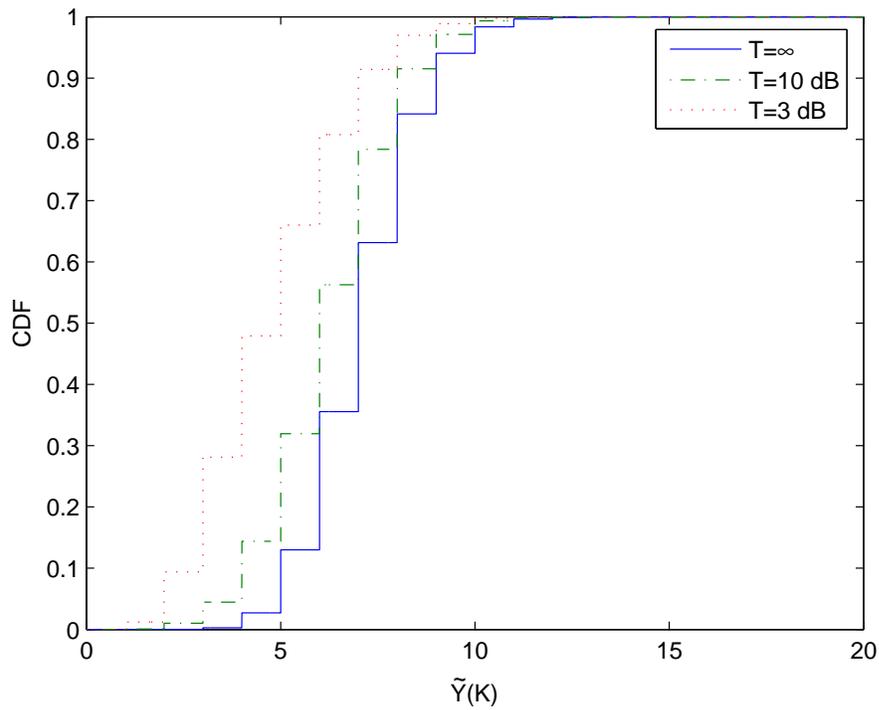}
   \caption{CDF of $\max_{j=1,\ldots, N_f} \tilde{\tilde{K}}_j^k$ for $N_f=5$ and various threshold values.}
   \label{fig3_2}
\end{figure}

\begin{figure}[ht]
   \includegraphics[scale=0.9]{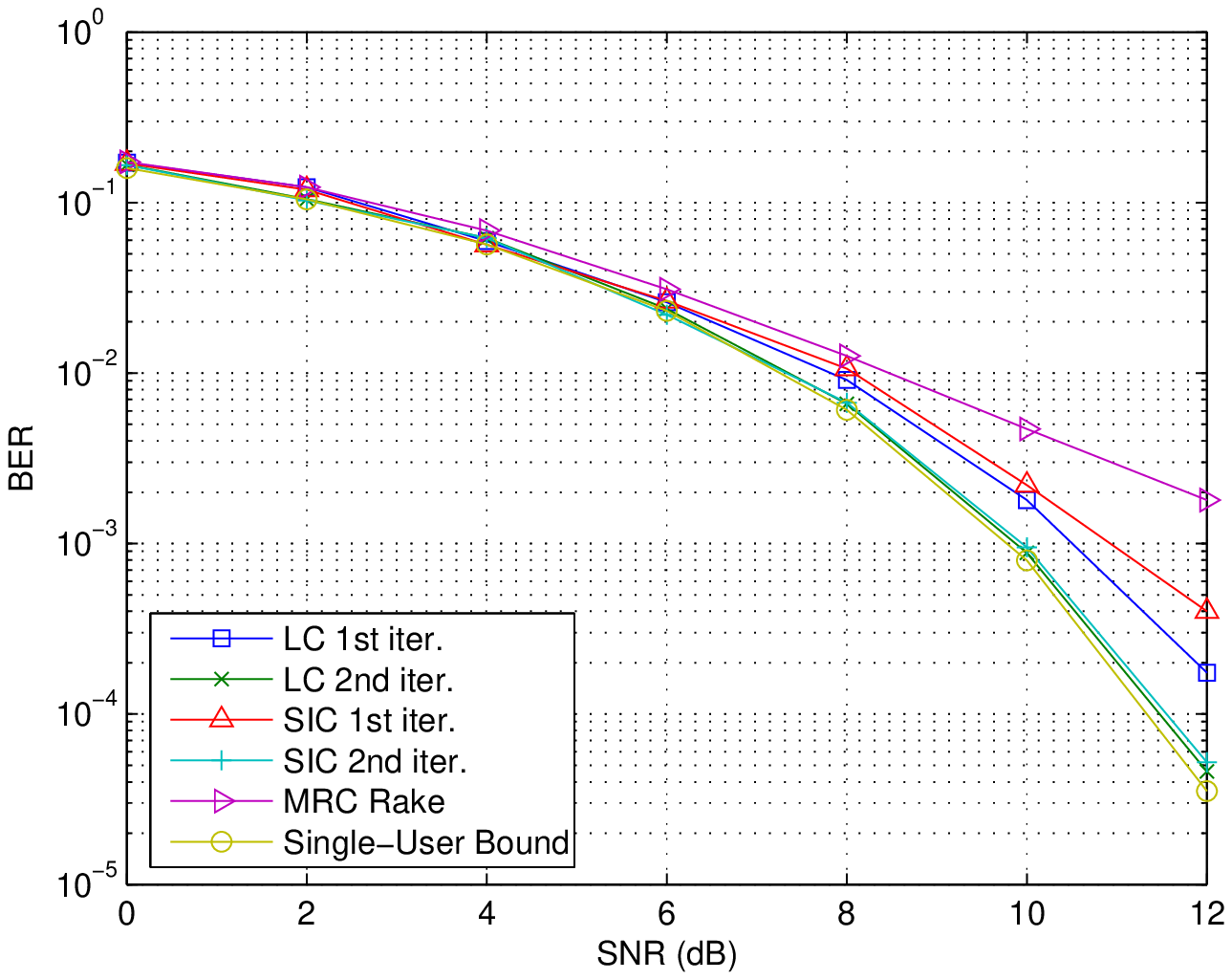}
   \caption{BER as a function of the SNR for various receivers.}
   \label{fig:BEP}
\end{figure}

\begin{figure}[ht]
   \includegraphics[scale=0.9]{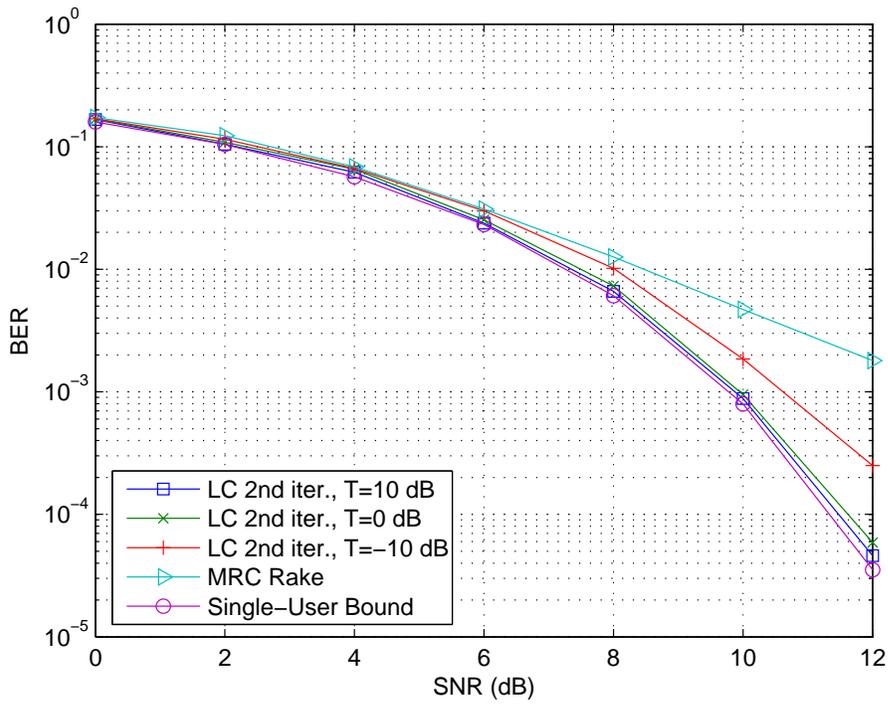}
   \caption{BER as a function of the SNR for various receivers, where the Gaussian approximation technique is plotted for various threshold values.}
   \label{fig:BEP_thr}
\end{figure}

\begin{figure}[ht]
   \includegraphics[scale=0.9]{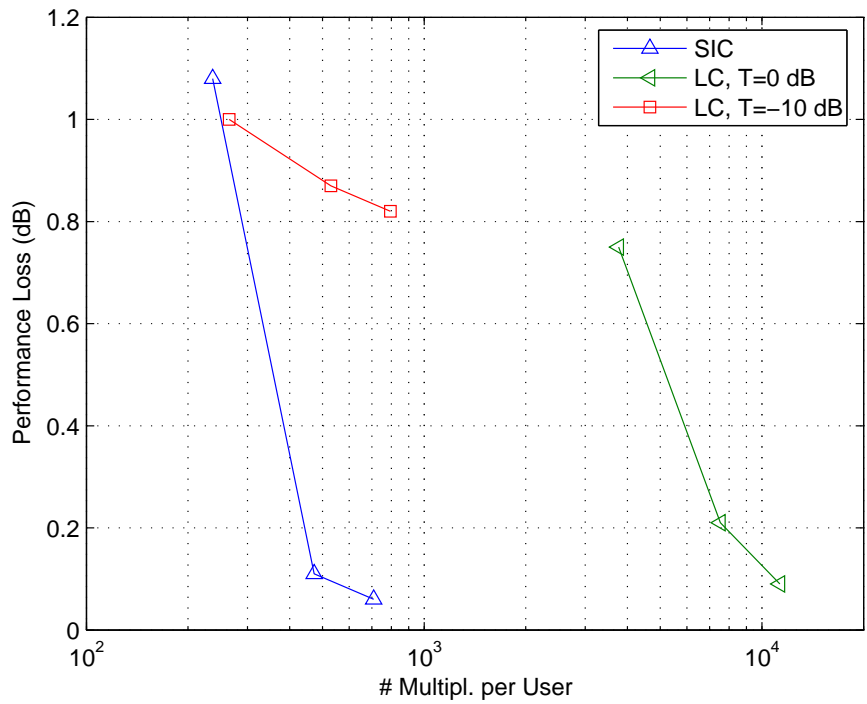}
   \caption{The distance in dB from a single user system at BER=$10^{-3}$ versus the average number of multiplication operations per user. For each receiver, the points are obtained for $1$, $2$ and $3$ iterations (${\Lcal}^1=\{1,\ldots,10\}$).}
   \label{fig:BEP_comp}
\end{figure}

\begin{figure}[ht]
   \includegraphics[scale=0.9]{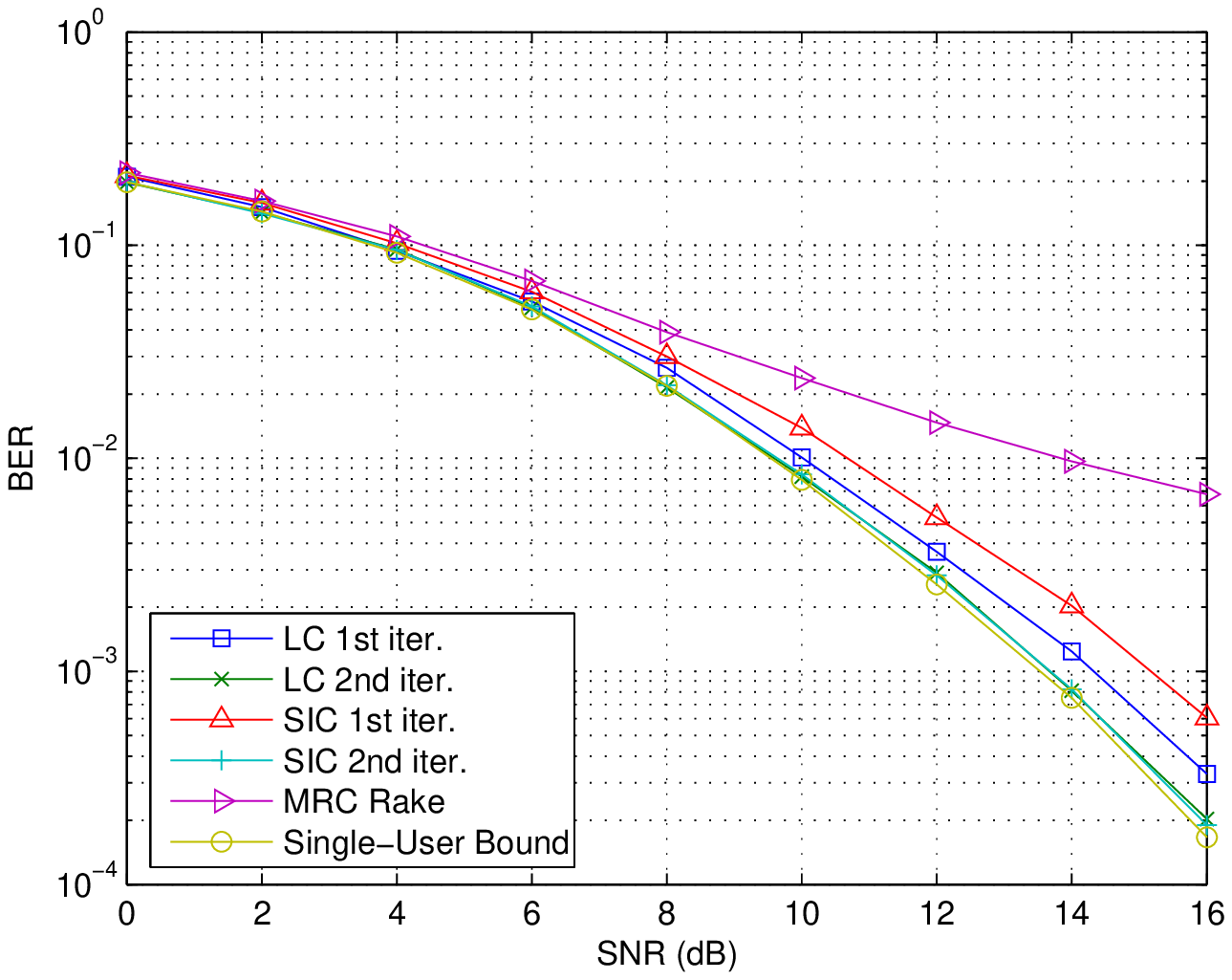}
   \caption{BER as a function of the SNR for various receivers.}
   \label{fig:BEP_5}
\end{figure}

\end{document}